\documentclass[prb,preprint]
{revtex4-1} 

\usepackage{amsmath}  
\usepackage{amsfonts} 
\usepackage{graphicx} 

\begin{document}


\title{Calculation of magnetic dipole moment of torus knots }

\author{J\'ulio Nascimento}
\email{julio.cesar.do.nascimento@hotmail.com} 
\affiliation{Departmento de F\'{\i}sica, Universidade Federal Rural de Pernambuco \\ 52171-900, Recife, PE, Brazil}

\author{Fernando Moraes}
\email{fernando.jsmoraes@ufrpe.br}
\affiliation{Departmento de F\'{\i}sica, Universidade Federal Rural de Pernambuco \\ 52171-900, Recife, PE, Brazil}


\date{\today}

\begin{abstract}
The magnetic moment of a closed loop with a steady current distribution is thoroughly described in many classical physics textbooks. Although usually the examples studied assume plane loops for explicit calculations of the magnetic moment, many other loop forms can be calculated. Among them, there are the torus knots. In this work, we explain how to calculate the magnetic moment of such current arrangement, and how it can be used as an interesting example that, even on three-dimensional loops, the magnetic moment can be taken as a simple product of the current and the vectorial area of the loop.
\end{abstract}

\maketitle 

\section{Introduction} 

In order to present an alternative to a steady current distribution that is not contained in a plane, we turn our eyes to the torus knot. Such arrangement is well known in the field of knot theory, that began with a series of papers published in 1877 by P. G. Tait, addressing the enumeration of knots. The motivation for such study was Lord Kelvin's theory of the atom. The theory stated that chemical properties of the elements were related to knotting that occurs between atoms, implying that insights into chemistry would be gained with an understanding of knots\cite{livingston1993knot}. 

The  importance  and  prevalence  of  knots  have  become  truly  apparent  and  this  has  
attracted increasing interest from scientists in different fields.  
In nature, molecular knots (including slipknots and pseudo-knots) are found throughout biology and exist in three major 
classes  of  biopolymers:  DNA,  RNA  and  proteins \cite{Sumners95,Matthews2010,Orlandini2017,Orlandini2018}. Molecular knots are increasingly   becoming   targets   of   chemical   synthesis.\cite{0953-8984-27-35-354101,Forgan2011}. References to knots can also be found in fundamental areas of physics, such as particle physics\cite{Faddeev1996,Witten1989} and electromagnetism \cite{Hirshfeld1998,Cantarella2001}.

One of the ways to characterize the interaction of systems with external fields is through its magnetic dipole \cite{Levin1984,Goedecke1999,Corbo2009}. Some models have been proposed to mathematically define these dipoles, mainly due to the fact that the magnetic dipoles are the most basic structure of  magnetism, considering the non-existence of magnetic monopoles\cite{Tellegen1962,Boyer1988,Fisher1971,Rosser1993}. Taking these considerations into account, and realizing that dipole moments play an important role in the training of a physicist, we are able to connect  knots and dipole calculations, providing an option to the dipole calculation of a non-planar loop. 

A knot means  a self-entangled closed loop that cannot be unraveled except by cutting the loop. The torus knot is a particular knot geometry, satisfying the condition that they all lie on a donut-shaped surface (i.e., a torus). The class of a torus knot is uniquely identified by a pair of relatively prime integers p and q; \cite{cromwell_2004} If p and q are not co-prime, then we have a collection of two or more identical knots that all lie on the same torus. The lower right panel of Fig. 1 shows schematically the way a (2,3) torus knot wraps around an underlying donut-shaped surface. The  integer p specifies the number of times the curve wraps around the rotational axis of the donut, while q specifies the number of times the curve passes through the hole of the donut, as visually depicted in Fig. 2. In terms of Cartesian coordinates, the (p,q)-torus knot is parameterized by \cite{PhysRevB.86.035415}:

\begin{figure}[h!]
\centering
\includegraphics[scale=.35]{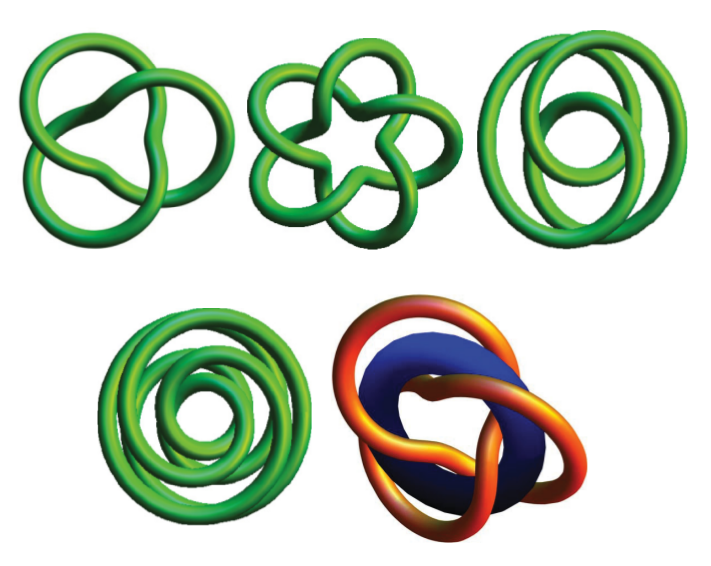}
\caption{Diagram of the four simplest torus knots
labeled by (p,q) = (2,3),(2,5),(3,2),(5,2), respectively. The lower
right panel illustrates how the (2,3) torus knot wraps around the
underlying torus colored in blue.\cite{PhysRevB.86.035415}}
\label{gasbulbdata}
\end{figure}

\begin{eqnarray}
x(t) &=& [R + \varepsilon cos(qt)]cos(pt),\nonumber \\
y(t) &=& [R + \varepsilon cos(qt)]sin(pt),\\
z(t) &=& \varepsilon sin(qt).\nonumber \\
\nonumber
\label{parametrization1}
\end{eqnarray}
with a variable t ($0\leq t \leq 2\pi$). The two constants $R$ and $\varepsilon$ ($0 < \varepsilon < R$) are called the major and minor radii of the torus, respectively; $\varepsilon$ is the radius of the tube of which the torus consists, and $R$ is the radius of the circle made of the tubular axis. In the following, we assume that $p,q > 0$ without loss of generality; this is because the $(p,-q)$ torus knot is the mirror image of the $(p,q)$ torus knot and the $(-p,-q)$ torus knot is equivalent to the $(p,q)$ torus knot except for the reversed orientation.

\section{Magnetic moment calculation}

We use the vector potential method to find the magnetic field of a small loop of current. By small we mean simply that we are interested in the fields only at distances large compared with the size of the loop. In making the multipole expansion of the vector-potential, it will turn out that any small enough loop can be approximated by a magnetic dipole.

\begin{figure}[h!]
\centering
\includegraphics[scale=.5]{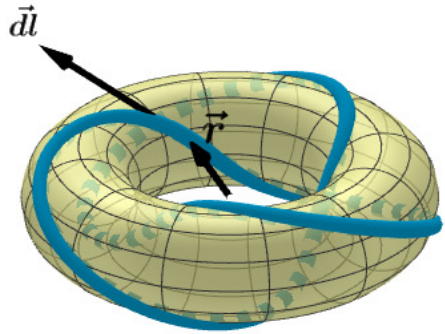}
\caption{Depiction of the trefoil that is the torus knot with $p=2$ e $q=3$, with the position vector $\vec{r}$ and the tangential vector $\vec{dl}$.}
\label{torus_completo}
\end{figure}

This magnetic dipole vector potential can be written in terms of a quantity called magnetic moment, that can be expressed in the SI units system by\cite{jackson_classical_1999} 
\begin{equation}
\vec{m}=\frac{I}{2}\int \vec{r} \times d\vec{l} ,
\label{magMomentint}
\end{equation}
where $I$ is the electric current circulating in the loop.
To determine $\vec{m}$ we have to define $\vec{r}$ and $d\vec{l}$ in the case of the torus knot, which is done in Fig. \ref{torus_completo}.

Visibly the position vector $\vec{r}$ is directly identified by the parameterization defined in Eq. (\ref{parametrization1}). The infinitesimal displacement in the metric space, can be defined as $d\vec{l}=dx\hat{x} + dy\hat{y} + dz\hat{z}$, that can be written with respect to the parametric variable $t$ as
\begin{equation}
d\vec{l} = \left(\frac{dx(t)}{dt}dt\right)\hat{x} + \left(\frac{dy(t)}{dt}dt\right)\hat{y} + \left(\frac{dz(t)}{dt}dt\right)\hat{z} .
\end{equation}
Taking this into a account we are able to write:

\begin{equation}
\begin{aligned}
\vec{r} \times d\vec{l} = \left[ y(t)\left(\frac{dz(t)}{dt}dt\right) - z(t)\left(\frac{dy(t)}{dt}dt\right)\right]\hat{x} + \\ 
\left[ z(t)\left(\frac{dx(t)}{dt}dt\right) - x(t)\left(\frac{dz(t)}{dt}dt\right)\right]\hat{y} +\\ 
\left[ x(t)\left(\frac{dy(t)}{dt}dt\right) - y(t)\left(\frac{dx(t)}{dt}dt\right)\right]\hat{z}.\hspace{0.3cm} 
\end{aligned}
\label{vectorProd}
\end{equation}
 
Eq. (\ref{vectorProd}) is a generalization, that can be used to compute the magnetic moment of any smooth parametric curve. Substituting this result in Eq. (\ref{magMomentint}), we get

\begin{equation}
\begin{aligned}
\vec{m} = \int_0^{2\pi} \left[ y(t)\left(\frac{dz(t)}{dt}\right) - z(t)\left(\frac{dy(t)}{dt}\right)\right]dt\quad \hat{x} + \\ 
\int_0^{2\pi}  \left[ z(t)\left(\frac{dx(t)}{dt}\right) - x(t)\left(\frac{dz(t)}{dt}\right)\right]dt \quad \hat{y} +\\ 
\int_0^{2\pi}  \left[ x(t)\left(\frac{dy(t)}{dt}\right) - y(t)\left(\frac{dx(t)}{dt}\right)\right]dt\quad \hat{z}.\hspace{0.3cm} 
\end{aligned}
\label{vectorProdIntegral}
\end{equation}
Now, using the torus knot parametrization, one can calculate the magnetic moment of the torus knot arrangement, that is expressed by:

\begin{equation}
\vec{m}=\frac{1}{2}I(\pi p\epsilon ^2 + 2\pi p R^2)\hat{z}.
\label{magMomentTorus}
\end{equation}

The formula in Eq.(\ref{magMomentTorus}) describes how the knotted geometry will be affected by an external magnetic field. Two main aspects need discussion, the first one is that due to the symmetry of the knot, the magnetic moment  has non-zero component only in the direction of higher symmetry. The second one is the fact that it is independent of $q$ 

From this result, we are able to discuss an important characteristic. The fact that the cross product in Eq. (\ref{magMomentint}) forms the area of the loop when integrated. When the discussed loop is planar, the vectorial area of the loop is usually simple to determine.
To make our point, let us define the area of the orthogonal projection of the knot onto the $x-y$ plane as
\begin{equation}
\vec{A}=\int x(t)\frac{dy(t)}{dt}dt .
\label{magMoment}
\end{equation}
Plugging into this formula the parametric curve of the torus knot, we get the area of the torus knot as seen from the z axis, which results in:
\begin{equation}
\vec{A}=\frac{1}{2}(\pi p\epsilon^2 + 2\pi p R^2)\hat{z}
\label{areaTorus}
\end{equation}

\begin{figure}[h!]
\centering
\includegraphics[scale=.6]{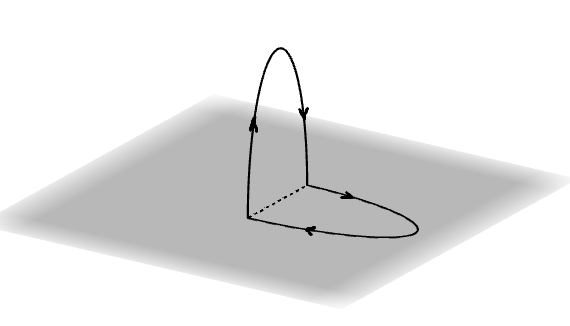}
\caption{Proposed system: bent elliptical loop.}
\label{loop1}
\end{figure}

\begin{figure}[h!]
\centering
\includegraphics[scale=.6]{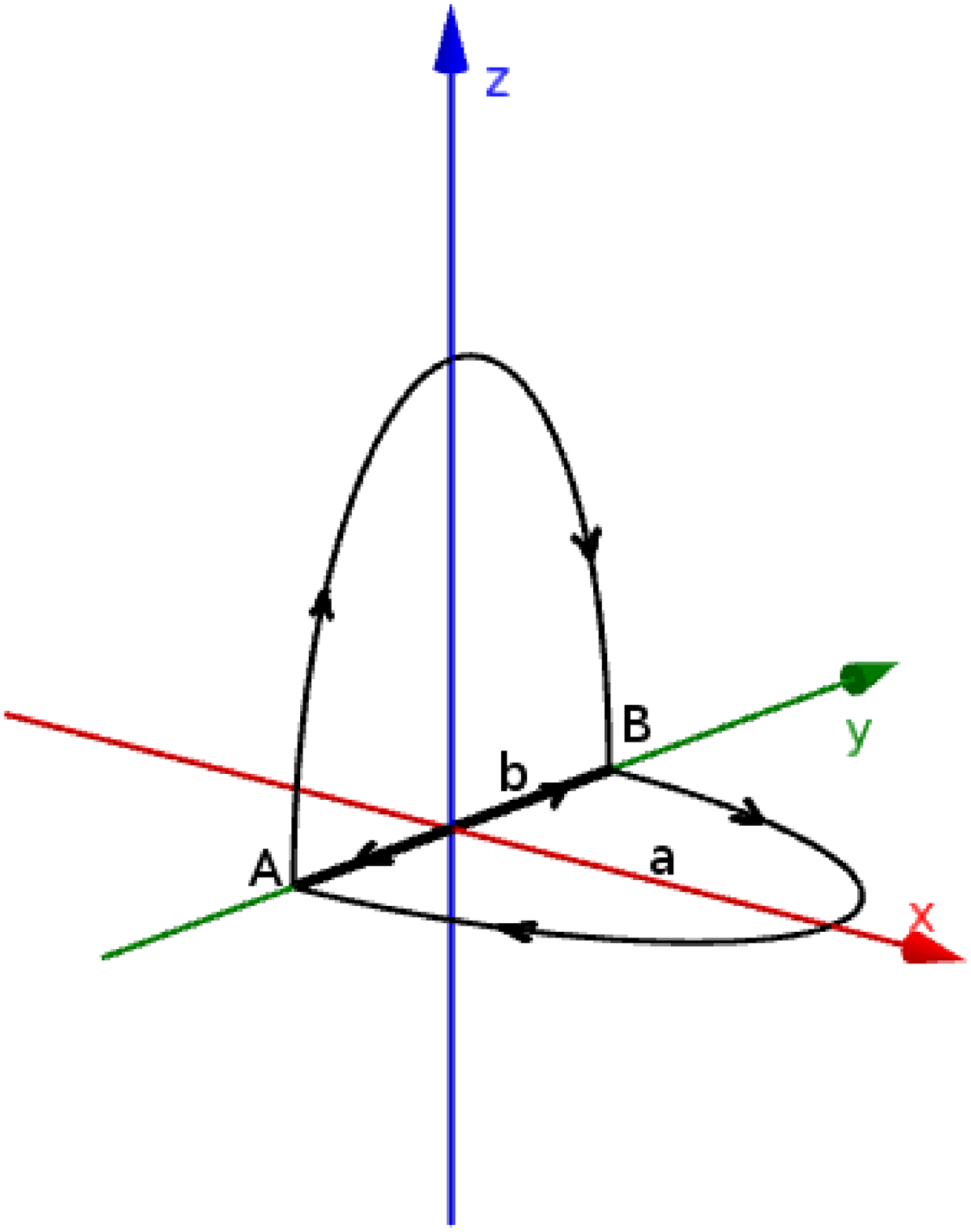}
\caption{Two half ellipses with major axis $a$ and minor axis $b$, connected by the line segment AB loop together to make up a system that has the same physical properties than the one proposed in Fig. \ref{loop1}}
\label{loop2}
\end{figure}

The similarity of Eq. (\ref{magMomentTorus}) and Eq. (\ref{areaTorus}) is the main point of this work. We are able to see that, for such system, the vectorial area of the knot can be taken as the projection that the knot makes onto the plane perpendicular to its direction of higher symmetry, in this case the $\hat{z}$ direction.     
Taken this into account, we are able to discuss the problem of an elliptical loop that has a static current. The loop has been folded as shown in the Fig. \ref{loop1}. This problem can be viewed as two half ellipses in  perpendicular planes from each other, each with a current with the same intensity as shown in Fig. \ref{loop2}.

\noindent
Note that the  the AB segment, at the intersection of the half ellipses, has currents with the same intensity, but in opposing directions, making the total current in the intersection zero, which turns this system equivalent to the one proposed in Fig. \ref{loop1}.

This way the problem is reduced to find the magnetic moment of two half ellipses, that can be obtained from the product of the area of each loop and the corresponding current, since both of them are planar loops. Considering that the major axis of the ellipses is taken as $a$ and the minor as $b$, from the  coordinate system proposed in the Fig. \ref{loop2} we have:

\begin{equation}
\vec{m}=\frac{I\pi ab}{2}(\hat{x} + \hat{z}).
\label{magMoment22}
\end{equation}

\begin{figure}[h!]
\centering
\includegraphics[scale=.6]{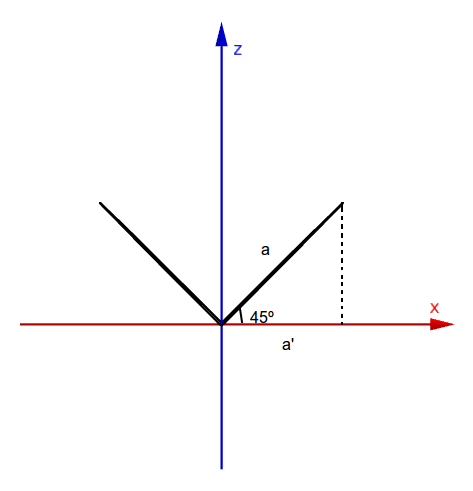}
\caption{Loop rotaded 45 degrees counter-clockwise around the $\hat{y}$ axis.}
\label{loop3}
\end{figure}

Another way to obtain this magnetic moment is to apply the same idea of the projection, used for the trefoil. In order to accomplish this  we  rotate the bent loop so that the axis of higher symmetry coincides with the z axis. To achieve that, let us rotate the loop 45 degrees counter-clockwise around the $\hat{y}$ axis. This way we get the configuration shown in Fig. \ref{loop3}, that is the view of the loop as seen from the $\hat{y}$ axis.

Comparing the projection of the folded loop with the original ellipse we realize the the minor axis does not change, but the major axis is altered becoming $a' = a cos(\pi/4)$. Using this we are able to calculate the magnetic moment as
\begin{equation}
\vec{m}=I\pi a'b \hat{z} = I\pi a\frac{\sqrt{2}}{2}b \hat{z}.
\label{magMoment3}
\end{equation}
If we now rotate back the loop, and so also rotate back the magnetic moment, we get:

\begin{equation}
\begin{pmatrix}
cos(\pi/4) & 0 & sin(\pi/4) \\ 
0 & 0 & 0\\ 
sin(\pi/4) & 0 & cos(\pi/4) 
\end{pmatrix}{\begin{pmatrix}
0\\ 
0\\ 
m
\end{pmatrix}} = {\begin{pmatrix}
msin(\pi/4)\\ 
0\\ 
mcos(\pi/4)
\end{pmatrix}} = {\begin{pmatrix}
\frac{m\sqrt{2}}{2}\\ 
0\\ 
\frac{m\sqrt{2}}{2}
\end{pmatrix}}.
\end{equation}

In vector form, it yields:

\begin{equation}
\vec{m}=\frac{I\pi ab}{2}(\hat{x} + \hat{z}),
\label{magMoment4}
\end{equation}
that is the exact same result as we obtained in Eq. (\ref{magMoment22}).

\section{Conclusion}

Taking in account the given results, we realize that, for loops that extend outside of a plane and have an axis of symmetry, the magnetic moment can be calculated as the product of the current and the area of the projection of the loop onto the plane normal to its axis of symmetry. Such realization greatly simplifies the calculation of magnetic moments of non-planar loops. We hope that this result will be helpful as a pedagogical tool and for gaining a deeper understanding of systems with torus knot symmetry.

\begin{acknowledgments}

We acknowledge the financial support of CNPq, CAPES and FACEPE.

\end{acknowledgments}


\begin{thebibliography}{0}%
\makeatletter
\providecommand \@ifxundefined [1]{%
 \@ifx{#1\undefined}
}%
\providecommand \@ifnum [1]{%
 \ifnum #1\expandafter \@firstoftwo
 \else \expandafter \@secondoftwo
 \fi
}%
\providecommand \@ifx [1]{%
 \ifx #1\expandafter \@firstoftwo
 \else \expandafter \@secondoftwo
 \fi
}%
\providecommand \natexlab [1]{#1}%
\providecommand \enquote  [1]{``#1''}%
\providecommand \bibnamefont  [1]{#1}%
\providecommand \bibfnamefont [1]{#1}%
\providecommand \citenamefont [1]{#1}%
\providecommand \href@noop [0]{\@secondoftwo}%
\providecommand \href [0]{\begingroup \@sanitize@url \@href}%
\providecommand \@href[1]{\@@startlink{#1}\@@href}%
\providecommand \@@href[1]{\endgroup#1\@@endlink}%
\providecommand \@sanitize@url [0]{\catcode `\\12\catcode `\$12\catcode
  `\&12\catcode `\#12\catcode `\^12\catcode `\_12\catcode `\%12\relax}%
\providecommand \@@startlink[1]{}%
\providecommand \@@endlink[0]{}%
\providecommand \url  [0]{\begingroup\@sanitize@url \@url }%
\providecommand \@url [1]{\endgroup\@href {#1}{\urlprefix }}%
\providecommand \urlprefix  [0]{URL }%
\providecommand \Eprint [0]{\href }%
\providecommand \doibase [0]{http://dx.doi.org/}%
\providecommand \selectlanguage [0]{\@gobble}%
\providecommand \bibinfo  [0]{\@secondoftwo}%
\providecommand \bibfield  [0]{\@secondoftwo}%
\providecommand \translation [1]{[#1]}%
\providecommand \BibitemOpen [0]{}%
\providecommand \bibitemStop [0]{}%
\providecommand \bibitemNoStop [0]{.\EOS\space}%
\providecommand \EOS [0]{\spacefactor3000\relax}%
\providecommand \BibitemShut  [1]{\csname bibitem#1\endcsname}%
\let\auto@bib@innerbib\@empty
\end{thebibliography}%


\begin{thebibliography}{99}

\bibitem{livingston1993knot}
C. Livingston.
\newblock {\em Knot Theory}.
\newblock Number v. 24 in Carus mathematical monographs. Mathematical Association of America, 1993.
 
\bibitem{Sumners95}
De Witt Sumners.
\newblock Lifting the curtain: using topology to probe the hidden action of enzymes.
\newblock {\em Notices of the AMS}, 42:528, 1995.

\bibitem{Matthews2010}
R. Matthews and A. A. Louis and J. M. Yeomans.
\newblock Effect of topology on dynamics of knots in polymers under tension.
\newblock {\em EPL (Europhysics Letters)}, 89:2, 2010.

\bibitem{Orlandini2017}
Enzo Orlandini and Guido Polles and Davide Marenduzzo and Cristian Micheletti.
\newblock Self-assembly of knots and links.
\newblock {\em Journal of Statistical Mechanics: Theory and Experiment}, 2017:3, 2017.

\bibitem{Orlandini2018}
Enzo Orlandini.
\newblock Statics and dynamics of DNA knotting.
\newblock {\em Journal of Physics A: Mathematical and Theoretical}, 51:5, 2018.

\bibitem{0953-8984-27-35-354101}
Nicole C~H Lim and Sophie~E Jackson.
\newblock Molecular knots in biology and chemistry.
\newblock {\em Journal of Physics: Condensed Matter}, 27(35):354101, 2015.

\bibitem{Forgan2011}
Forgan, Ross S. and Sauvage, Jean-Pierre and Stoddart, J. Fraser.
\newblock Chemical Topology: Complex Molecular Knots, Links, and Entanglements.
\newblock {\em Chemical Reviews}, 111(9):5434, 2011.

\bibitem{Faddeev1996}
Faddeev, L. D. and Niemi, Antti J..
\newblock Knots and particles.
\newblock {\em Nature}, 387:58, 1997.

\bibitem{Witten1989}
Witten, E.
\newblock Quantum field theory and the Jones polynomial.
\newblock {\em  E. Commun.Math. Phys.}, 121:351, 1989.

\bibitem{Hirshfeld1998}
Hirshfeld,Allen C.
\newblock Knots and physics: Old wine in new bottle.
\newblock {\em  American Journal of Physics}, 66(12):1060, 1998.

\bibitem{Cantarella2001}
Cantarella,Jason  and DeTurck,Dennis  and Gluck,Herman.
\newblock Knots and physics: Old wine in new bottle.
\newblock {\em  Journal of Mathematical Physics}, 42(2):876, 2001.

\bibitem{Levin1984}
Levin,Eugene 
\newblock Magnetic dipole moment measurement
\newblock {\em  American Journal of Physics}, 52(3):248, 1984.

\bibitem{Goedecke1999}
Goedecke,G. H.  and Wood,Roy C.  and Nachman,Paul
\newblock Magnetic dipole orientation energetics
\newblock {\em  American Journal of Physics}, 67(1):45, 1999.

\bibitem{Corbo2009}
Corbò,Guido  and Testa,Massimo 
\newblock Magnetic dipoles and electric currents
\newblock {\em  American Journal of Physics}, 77(9):818, 2009.

\bibitem{Tellegen1962}
Tellegen,B. D. H. 
\newblock Magnetic-Dipole Models
\newblock {\em  American Journal of Physics}, 30(9):650, 1962.

\bibitem{Boyer1988}
Boyer,Timothy H. 
\newblock The force on a magnetic dipole
\newblock {\em  American Journal of Physics}, 56(8):688, 1988.

\bibitem{Fisher1971}
Fisher,George P. 
\newblock The Electric Dipole Moment of a Moving Magnetic Dipole
\newblock {\em  American Journal of Physics}, 39(12):1528, 1971.

\bibitem{Rosser1993}
Rosser,W. G. V. 
\newblock Classical electromagnetism and relativity: A moving magnetic dipole
\newblock {\em  American Journal of Physics}, 61(4):371, 1993.

\bibitem{cromwell_2004}
Peter~R. Cromwell.
\newblock {\em Knots and Links}.
\newblock Cambridge University Press, 2004.

\bibitem{PhysRevB.86.035415}
Hiroyuki Shima.
\newblock Persistent current in quantum torus knots.
\newblock {\em Phys. Rev. B}, 86:035415, Jul 2012.

\bibitem{jackson_classical_1999}
John~David Jackson.
\newblock {\em Classical Electrodynamics}.
\newblock John Wiley \& Sons, 2012.

%

  


















\end{thebibliography}
\end{document}